\documentclass[12pt]{article}

\begin{document}

\title{Different scenarios of the late stages of condensation }

\author{Victor Kurasov}

\date{  }

\maketitle

\begin{abstract}
The late stages of the nucleation have been described
analytically. The approximate solution of the
Zel'dowich-Folmer-Frenkel equation has been constructed.
\end{abstract}

\section*{Introduction}

The process of the evolution after the end of nucleation (the
main stage) and the consumption of the main quantity of the
surplus metastable phase requires a separate description.
This description can be based on the real form of the size
spectrum \cite{preprint} or
can be done  by the asymptotic analysis
\cite{ls}, \cite{kukos}.

The regime will be chosen
as the free molecular one. This is done because the
formation of the exhausted zones around the droplets  strongly influences
the kinetics and the description can not be based on the
averaged characteristics. The only situation where the
diffusion regime can be combined with the averaged
characteristics is when the embryos go from one cave to
another and change many exhausted zones.
But this situation can be reduced to the effective absence
of the cave and, thus, the regime will be the free-molecular
one. It is important because
earlier there was the restriction that the
size is less than the length of the free motion of a
molecule divided by the coefficient
of condensation. Now this restriction is not essential.

The first stages of the late evolution were investigated in
\cite{preprint}.

\section{Balance equation}

Define $n$ as the number of embryos with $\nu$ molecules.
The evolution in time $t$ is described by
equation
$$
\frac{\partial n}{\partial t} = W^+(\nu-1) n(\nu-1)
-W^-(\nu) n(\nu)
-W^+(\nu) n(\nu)
+W^-(\nu+1) n(\nu+1)
$$
Here $W^+$ is the direct kinetic coefficient, $W^-$ is the
inverse kinetic coefficient. In the Fokker-Planck
approximation the will be a diffusion and the regular
growth.

In approximation of the pure diffusion the balance equation will be
$$
\frac{\partial n}{\partial t} = W^+(\nu)[ n(\nu-1)
- n(\nu)
- n(\nu)
+ n(\nu+1) ]
$$
and
$$
\frac{\partial n}{\partial t} = W^+(\nu)
\frac{\partial^2 n(\nu)}{\partial \nu^2}
$$

\section{Some estimates }

According to \cite{preprint} after the regular relaxation there
will be a diffusion errosion of the spectrum. For this
stage one can give some estimates.

\subsection{Diffusion}

One has to use $s$-scale. Here $s$ is the surface of the
embryo
$$
s = \nu^{2/3}
$$
 This is because the kinetic coefficient of
absorption $W^+$ is proportional to $s$
$$
W^+ = W^0 s
$$
 Then in diffusion
equation
will be
\begin{equation}\label{purdif}
\frac{\partial n}{\partial t} = W^0 \nu^{2/3}
\frac{\partial^2 n(\nu)}{\partial \nu^2}
\end{equation}

In the leading term
$$
\frac{\partial^2 n(\nu)}{\partial \nu^2}
\rightarrow
\nu^{-2/3}
\frac{\partial^2 n(\nu)}{\partial s^2}
$$

Then
$$
\frac{\partial n(\nu)}{\partial t} = W^0
\frac{\partial^2 n(\nu)}{\partial s^2}
$$
The same will be for $n(s)$.

The diffusion scale will be
$$
s_{diff} = t^{1/2}
$$
For the linear size of an embryo
$$
\rho = \nu^{1/3}
$$
we have
$$
\rho_{diff} \sim t^{1/4}
$$

\subsection{Regular growth}

The coordinate of the regular growth will be determined
from
$$
\frac{d \rho}{d t} = \frac{\zeta}{t_t}
$$
Here
$t_t$ is the characteristic time, $\zeta$ is the
supersaturation defined by
$$
\zeta = n_1/n_\infty - 1
$$
where $n_1$ is the molecules number density of vapor, $n_\infty$ is the
 molecules number
density of the saturated vapor.

Then
$$
\rho_{reg} = \int \frac{\zeta}{t_t} dt
$$

The behavior of $\zeta $ is given by
$$
\zeta = \frac{2a}{3\rho_c}
$$
where index $c$ marks the critical embryo.

Having taken
$$
\rho_c \sim \rho_{diff}
$$
we get
$$
\zeta \sim \frac{2a}{3\rho_c}|_{\rho_c = \rho_{diff}}
$$
Then
$$
\zeta \sim \frac{2a}{3\rho_c}|_{\rho_c = t^{1/4}}
$$
and
$$
\zeta \sim  t^{-1/4}
$$

Then
$$
\rho_{reg} \sim t^{3/4}
$$

We see that
$$
\rho_{diff} \ll \rho_{reg}
$$
and it is clear that the main quantity of substance will be
in the tail growing regularly.

Approximately we assume that at the period when the
tail will be essential the following approximation $W^0 \approx  const$
takes place.

\section{The form of the tail}

We determine the form of the tail. Imagine that at big
sizes there is a dominating regular growth, at small sizes
of the critical order there is a dominating diffusion. The
boundary will be marked by the index $b$.

Certainly, at the
very small sizes there is a regular dissolution. But this
region is simply negligible and can be expelled by the
transition of a zero point.

Then for $\rho>\rho_b$
$$
\rho - \rho_b = \int_{t'}^t \frac{\zeta}{t_t} dt
$$

Then the distribution function $f$ is given by
$$
f(\rho, t ) = f_b(\rho_b, t')
$$
To get $f_b$ one can use the diffusion approximation
according to \cite{book2}, \cite{preprint}
$$
f_b = \frac{A_0}{\sqrt{4W^0t'}} \exp(-\frac{(s_b-s_c)^2}{4
W^0 t'})
$$
In the absolutely stationary approximation
one can put the normalizing constant
$$A_0 = 1$$
Due to the dissolution this constant will be another and
this question will be solved later.

Now we consider $$\frac{(s_b-s_c)^2}{4
W^0 t'}$$
As an estimate one can take
$$
\rho_b - \rho_c = 2 \rho_c
$$

Then
$$
\frac{(s_b-s_c)^2}{4
W^0 t'}
\sim
\frac{64 \rho_c^4}{4 W^0 t'}
$$

If we take $\rho_c \sim t^{1/4}$ (more precise it is
necessary to take $\rho_c \sim W^0 t^{1/4}$) then
$$
\frac{(s_b-s_c)^2}{4
W^0 t'}
\sim
\frac{64 }{4}
$$
Then
$$
f_b \sim
\frac{A_0}{\sqrt{4W^0t'}} \exp(-16)
$$

\section{More precise formula}

One can use another more precise equation
(see \cite{book2}, \cite{preprint})
$$
f_b =
\frac{A_0}{\sqrt{4W^0t}} \exp(-\frac{(s-s_{c0})^2}{4
W^0 t})
-
\frac{A_0}{\sqrt{4W^0t}} \exp(-\frac{(s+s_{c0})^2}{4
W^0 t})
$$
Here $s_{c0}$ marks the initial position of a critical
surface.

After the transformations we get
$$
f =
\frac{A_0}{\sqrt{4W^0t}}
\exp(-\frac{s^2}{4
W^0 t})
\exp(-\frac{s_{c0}^2}{4
W^0 t})
[\exp(\frac{2 s s_{c0}
}{4
W^0 t})
-
\exp(- \frac{2 s s_{c0}
}{4
W^0 t})
]
$$

One can see that
$$
s s_{c0} / 4 W^0 t \ll 1
$$
Really, since
$$
s^2 \sim 4 W^0 t
$$ and $s_{co}$ is fixed
we come to
$$
s s_{c0} / 4 W^0 t \sim t^{-1/2} \ll 1
$$
Then
after the decompositions of the corresponding exponents one
can come to
$$
f =
\frac{A_0}{\sqrt{4W^0t}}
\exp(-\frac{s^2}{4
W^0 t})
\exp(-\frac{s_{c0}^2}{4
W^0 t})
\frac{ 4 s s_{co}}{4 W^0 t}
$$

So,
$$
f \sim y \exp(-y^2)
$$
with
$$
y = \frac{s}{\sqrt{4 W^0 t}}
$$

The maximum of this distribution is attained at
$$
y_m = \sqrt{2}/2
$$
This value will be close to the critical value
$$
y_m \sim y_c
$$

For
$$
y_b = \alpha y_c
$$
with some parameter $\alpha$ we get
$$
f_b = \frac{A_0}{\sqrt{4 W^0 t}} \exp(-\beta)
$$
with constant $\beta$.

\section{Dissolution  of the spectrum head}

The dissolution of the head determines the form of the
tail.

At first there is a pure dissolution of a gaussian and
$$
\rho_c \approx const
$$
Then
$$
f_b \sim \frac{1}{\sqrt{4 W^0 t}} \exp(-(s_b - s_c)^2 / (4
W^0 t ))
$$
or
$$
f_b \sim \frac{1}{\sqrt{4 W^0 t}} \exp(- 9 W^0 \rho_{c0}^4 / (4
W^0 t ))
$$
For the behavior of $f_b$  we get
$$
f_b \sim \frac{1}{t^{1/2}} \exp( -  \gamma / t)
$$
with a positive parameter $\gamma = const$.

The last function is positive, it has an evident asymptotic
behavior
$f_b \rightarrow t^{-1/2}$ at big $t$. At small $t$ it goes
to zero. Thus, it has a maximum.

To see the formation of
the tail one can approximately write
$$
\rho - \rho_{c0} = \frac{\zeta_0}{t_t} (t-t')
$$
where $\zeta_0$ is some initial supersaturation.
Here we assume $\zeta_0$ to be a constant which is a very
strong assumption. So, below all results will have the
power of stimates.

The maximal size $ \rho_{lim}$ is
$$
\rho_{lim} = \rho_{c0} + \zeta_0 t / t_t
$$

Then having put the starting moment at zero one can get
$$
t' = (\rho - \rho_{lim}) t_t/ \zeta_0
$$

The distribution at the tail looks like
$$
\tilde{f}_\rho = f(z) = (\frac{\zeta_0}{z t_t})^{1/2}
\exp(-\frac{\gamma \zeta_0 }{z t_t})
$$
as a function of
$$
z = \rho_{lim} - \rho
$$

Now we shall calculate the quantity of substance in the
tail $G_{tail}$. We get
$$
G_{tail}
=
\int_{3\rho_c}^{\rho_{lim}}
\rho^3
(\frac{\zeta_0}{t_t(\rho_{lim} - \rho)})^{1/2}
\exp(-\gamma \frac{\zeta_0}{t_t(\rho_{lim} - \rho)})
d \rho
$$

Having introduced
$$
y = \frac{\rho_{lim} - \rho}{\epsilon} \ \ \ \  \ \epsilon=
\gamma \zeta_0 / t_t
$$
we get
$$
G_{tail} =
\frac{\epsilon}{\gamma^{1/2}}
\int_0^{\frac{\rho_{lim} - 3 \rho_c}{t_t}}
(\rho_{lim} - \epsilon y)^3 \frac{1}{y^{1/2}} \exp(-1/y) dy
$$

One can see that $G_{tail}$ grows very fast. The
asymptotics
is
$$
G_{tail} \rightarrow \rho_{lim}^{5/2} \sim t^{5/2}
$$

But this behavior will take place only until
$$
s s_c \sim 4 W^0 t
$$
After this moment of time the boundary condition will be
another
$$
f|_{\rho_b} = A_0 \exp(-const)
$$
or at larger scales of time
$$
f|_{\rho_b} \sim t^{-1/2}
$$

\subsection*{Alternative approach}

At the further stage
$$
f|_{\rho_b} \sim \frac{1}{\sqrt{t}}
$$

So, we see that the variation is very small. The drift of $\rho$ is
rather small also
$$
\rho \sim \int \zeta dt \sim t^{3/4}
$$
because of
$$
s_c \sim t^{1/2} \ \ \ \ \rho_c \sim t^{1/4} \  \ \ \
\zeta \sim t^{-1/4}
$$

The subintegral function $g_{tail}$ in the
quantity of substance
$$ G_{tail} = \int g_{tail} d \rho
$$
in the tail grows like
$$
g_{tail} \sim t^{9/4}
$$
It grows fast.

\section{Impulse regime}

Since $G_{tail}$ grows very fast  one gas to analyze this
behavior.

At first
\begin{equation}\label{p}
G_{tail} \ll G_{total}
\end{equation}
where $G_{total}$ is the total quantity of substance in
droplets.

The main role here is played by diffusion and the solution  can be
described with the help of combinations of Gaussians. This
model will be called as the diffusion-regular model. It is
investigated in \cite{preprint}.

Since
$g_{tail}$ grows fast it means that the monodisperse
approximation for $G_{tail}$ is quite suitable
$$
 G_{tail} \sim \rho_{eff} N_{eff}
 $$
 Here $N_{eff}$ is the effective number of droplets
 in the tail and $\rho_{eff}$ is their coordinate.
Certainly, $N_{eff}$ is not a constant value, it grows.

At $t$ determined by
$$
G_{tail} \sim G_{total} - G_{tail}
$$
the tail begins to eat the head of the spectrum.
Here the tail is non-essentially decreasing one.
It occurs
very fast and the spectrum eats at first the head and then
it begins to eat the the rather flat beginning of the tail.

This process can be easy described by the regular
growth of the tail and by the regular dissolution of the
eaten part of the spectrum.

Then only the end of the tail will exist. This end can be
then treated as a monodisperse spectrum.
Later this part will be   dissolved by diffusion and the process
repeats.

We shall call this process as the "impulse condensation".

This part is described in \cite{preprint}.

\section{Short tail}

A question whether the process of impulse
condensation repeats  arises here. From the first
point of view it seems that this process will infinitely
repeat. But there is one objection.

The function $f(t) = \exp(-\gamma/t)/\sqrt{t}$ described above has
a maximum. This maximum takes place at $t/\gamma \equiv y = y_m \approx 2$ and
for $y$ essentially less than $y_m$ there is a rapid
decrease to zero.

Certainly, after every cycle the time $t$ has to be
shifted.

The crucial point is whether at $y_m$ the condition
(\ref{p}) is observed. An alternative case
$$
G_{tail} \sim G_{total} - G_{tail}
$$
at $y_m$ leads to another further evolution.
Here the tail is essentially decreasing.
This case will
be called as "the case of adjusted tail".

In this section one can study the evolution on the base of
the regular growth. This radically simplifies the
situation.

Here we shall give the main ideas of description in this
situation.
The vapor consumption by the head of the tail can  eat only
the same head of the tail. There is no other parts to eat.
Then we come to the following balanced picture:
\begin{itemize}
\item
The task is to determine $\rho_b = 2\div3 \rho_c $,
$\rho_c = 2a/3 \ln(\zeta+1) \approx 2a/3\zeta$ where $a$ is
a renormalized surface tension.
\item
The size distribution for $\rho > \rho_b $ is known.
We have
$$ f(\rho, t) = f(x) $$ for $$x = z-\rho$$ $z\equiv
\rho_{lim}(t)$. Then
$$
G_{tail} = \int^{z-\rho_b}_0 (z-x)^3 f(x) dx
$$
\item
For $f(x)$ we have
$$
f(x) \sim \exp(-\frac{\tilde{\gamma}}{x}) / \sqrt{x}
$$
\item
The balance equation is
$$
G_{tail} + G_{head} + n_{\infty}(\zeta+1) = const
$$
where $G_{head} $ is the quantity of the substance in a
 region $0.7 \rho_c < \rho < \rho_b$. In the region
$\rho < 0.7 \rho_c$ one can simply neglect the substance.
\item
In the region $0.7 \rho_c < \rho < \rho_b$ one can write
the pure\footnote{I.e. without the regular growth and with the
constant coefficient of diffusion.} diffusion equation
(\ref{purdif}) with boundary conditions
$$f(\rho,t) = 0$$
at $\rho=0.7 \rho_c$
$$f(\rho,t) = f_{tail}(x)$$
at $\rho = \rho_b$
\item
In the region $\rho < 0.7 \rho_c$ one can simply neglect the
evolution.
\end{itemize}

\section{Further simplification}

Already the last system of equations can be solved when we
know $f(x)$. The function $f(x)$ has to be known from the
initial conditions. These conditions are:
\begin{itemize}
\item
The direct result of the nucleation period \cite{kuni}.
\item
The result of the impulse regime \cite{preprint}.
\end{itemize}

To show the result in a most simple way one can use some
simplifications.
Further simplification is the following
\begin{itemize}
\item
To write the balance equation as
$$
G_{tail} = fixed
$$
\item
To use that
$$
G_{tail} \sim N_{tail} \rho_b
$$
where $N_{tail}$ is the effective (proportional)
number of the droplets in the tail.
\end{itemize}

The function $f_{tail}$ is one of the possible
realizations, which has a certain disadvantage. Really, for
$\rho > \rho_{lim}$ the distribution is zero. Meanwhile,
the diffusion in the region $\rho> \rho_b$ inevitably leads
to the big droplets. Assume that the behavior of the gaussian at big
arguments can be approximated as
$$
\tilde{f} \sim  \exp ( - \gamma' \rho)
$$
with some parameter $\gamma'$.

The same asymptotic can be derived from $f_{tail}$  by the
steepest descent method.

So, the last approximation is very fruitful. It allows to
calculate the integrals for $G_{tail}$ analytically.

One can make some new simplification. Since the spectrum is
rather sharp, it means that
$$
\frac{ d \rho_b}{dt} \sim \frac{ d \rho_c}{dt}
$$
which leads asymptotically to
$$
\zeta \sim t^{-1/2}
$$
$$
\rho \sim t^{1/2}
$$
Since the diffusion width is $s_d \sim t^{1/2}$, $\rho_d
\sim t^{1/4}$ we see that the diffusion does not radically
change the character of the process. However, since the
Gaussian does not have a finite support
 the droplets with the
greatest sizes which will be the main in further
consumption appears due to diffusion. This is one more
point for the approximation $\tilde{f}$.

The constant $\gamma$ is not a precise
constant but a slowly varying function. This
effect can be taken into account by standard methods.

An approximation $\tilde{f}$ is so simple that we have no
need to use the monodisperse approximation.
One can calculate $G_{tail}$ directly by integration
$$
G_{tail} \sim  \int_{\rho_b}^{\infty} (z-x)^3 \exp(-\gamma' x) dx
$$
and the integral can be taken analytically.

\section{Initial asymptotic solution}

Since we came to the balance equations typical for the
asymptotic solutions of Lifshic and Slyozov \cite{ls} it is
necessary to analyze this approach.

As a regular law of droplets growth one has to take the
precise expression
$$
\frac{d \rho}{dt}
=
\frac{6 \sigma}{9 t_t} (\frac{1}{\rho_c} - \frac{1}{\rho})
$$
where $\sigma$ is the renormalized surface tension.

For $$
u= \rho / \rho_c
$$
one  gets
$$
\frac{du}{dt} =\frac{6 \sigma}{9 t_t
\rho_c } (\frac{1}{\rho_c} - \frac{1}{\rho}) -
\frac{u}{\rho_c} \frac{d\rho_c}{dt}
$$

In the reduced coordinates
$$
\frac{du}{d\tau} = (1-1/u) - \tilde{\gamma} u
$$
$$
\tilde{\gamma} = \rho_c \frac{d\rho_c}{dt}
$$

Then to escape the violence of the substance balance one
has to observe
$$
\frac{du}{dt} \leq 0
$$
for all $u$.

The last relation will take place for every time $\tau$
increasing in time since
$$
\frac{du}{d\tau} = \frac{du}{dt}\frac{dt}{d\tau}
$$
Namely
$$\tau = 3 \ln(\rho_c/\rho_c(t=0))$$ was chosen by
Lifshic, Slyozov (LS).

The last requirement
is very strong. In reality, one can imagine the
situation where this requirement is not valid. In the general case it is
necessary that this requirement takes place only in the
integral sense.
But  in LS  theory this requirement has to be valid at
every moment.

LS required that
$$max_{\{ u \}} (\frac{du}{d\tau}) \equiv r \rightarrow 0
$$
Meanwhile the
real condition should be the following:
\begin{itemize}
\item
If  $r$ has a power asymptotics (asymptotics as an argument in some power)
then this asymptotics
has to be
zero
\end{itemize}
But the situation when $r$ has no asymptotics is also quite possible.
For example, the case of oscillations drops out of
attention here.

 The crucial supposition of LS theory is the existence of
 the asymptotics for $r$ (in the class of constants).

The recipe of LS
is to consider $r$ as the monotonous function and then it
is reasonable to put $r=0$ asymptotically.

It is necessary to stress that the condition
$du/dt \leq 0$ for every $u$
leads to
$$
\frac{d\rho}{dt} - \frac{\rho}{\rho_c} \frac{d \rho}{dt}
\leq 0
$$ for every $\rho$
or
$$
\frac{d \ln \rho}{dt} \leq \frac{d \ln \rho_c}{dt}
$$ for every $\rho$.
This is very important relation.

From the last inequality it follows that every $\rho$ can
not escape and will inevitably be dissolved.
But at some $u$ there will be no dissolution. Let it be
$u_0$. Requirement $r=0$
leads to
$$
u_0 =2
$$
$$
\frac{
\frac{6 \sigma}{9 t_t}
}{\rho_c^2 \frac{d\rho_c}{dt}} = 4
$$

The last equation  can be easily integrated which leads to
\begin{equation}
\label{4/9}
-4 \rho_c^{-1} = const + t \frac{6 \sigma}{9 t_t}
\end{equation}
but one has to note that if we observe the validity of the
previous equation only in the averaged sense we get
approximately the same integral law. So, one can not regard
the law of $4/9$ (this is the characteristic coefficient in
the diffusion case, here the regime is the free-molecular one)
as an experimental justification of LS
approach.

The law (\ref{4/9}) can be easily justified. Really, if
there will be another power asymptotic then immediately we
come to the violence of the balance condition.

One has also to stress that it is absolutely impossible to
differentiate (\ref{4/9}), the result will have nothing in
common with a real situation. This fact is typical for such
asymptotics.

\section{The form of the spectrum in LS theory}

It is clear that the evolution of the system is governed by
the spectrum of the droplets sizes. This function
determines the behavior of the supersaturation in the
system.

In frames of LS approach the behavior of $\rho_c$
 is already  known.
Then it is possible to solve the inverse problem: to
reconstruct the spectrum on the base of the behavior of
supersaturation. The sense of this way of derivation is
wrong, but technically it is quite possible.

Under the regular law of growth one can find for the
distribution function
$$
\phi(\tau, u) =  - \xi(\tau - \tau(u))/v_u
$$
where
$$
v_u = du/d\tau
$$
is already known
and
$$
\tau(u)
= \int_0^u du/v_u
$$
is also known and
$\xi$ is some arbitrary function which is going to be
determined.
The balance equation allows to determine the form of $\xi$
which solves the problem.

To solve it in  the simple form $\tau$ is chosen as
$$\tau = 3 \ln(\rho_c(t)/\rho_c(t=0))$$. Then the balance
equation in a closed system will be
$$
\exp(\tau) \int_0^u u^3 \phi(\tau, u) du = 1
$$
which can take place only if
$$
\xi (...) = \exp(...)
$$

The spectrum is zero for $u>u_0$ and continuously goes to
zero for $u \rightarrow u_0-0$. So, the essential part of
the spectrum ill be dissolved at finite times $\tau$ and
$t$.
Here it is necessary to stress that now it is clear that
the deviation of $r$ from zero at some finite interval (it
is quite possible) can lead to the absolutely another form
of the spectrum. So, the form of the spectrum is determined
only with the help of the strong supposition of LS theory
about the behavior of the critical size, i.e. about the
supersaturation.

\section{Correction for $\gamma$}

The limit value $r\equiv 0 $ is not acceptable
even in LS theory. The next step leads to correction of
$r$.
Ordinary,
the correction is many times greater then the precedent
value of $r$ which allows Osipov and Kukushkin \cite{kukos}
to speak about the non-uniform  character of LS
decompositions.

 In frames of LS theory it is possible to write a
correction term and to fulfill the next step of
calculations. Osipov and Kukushkin (OK) \cite{kukos} follow
another way. Certainly, since the result in the first
approximation is known it is possible to rewrite the formulas
in such a way that $\gamma$ will be zero and the first
approximation has to inevitably include the correction term in
LS approach.  Since  the conclusions of OK theory are
important it is necessary to consider it.
We omit all details which can be found\footnote{The
initial total number of embryos is found with an error from \cite{kukos1}
because the asymptotic analysis \cite{kukos1} is not valid, it is necessary to use
\cite{kuni}.} in
\cite{kukos}.

Having introduced
$$\tau = \frac{1}{4} \int_0^t \rho_c^{-2} dt
$$
and redefined $u=\rho/2\rho_c$ one can come to
$$
\frac{du}{d\tau} =
\frac{(u-1)^2}{u} + \gamma u
$$
Now the condition $r=0$ corresponds to $\gamma=0$ and
every correction for $\gamma$ will be giant in comparison with $\gamma_0 =0$.
Then the
asymptotics here has to include the correction term already
in the main order.

The  approach with correction gives $$\gamma = 1/(4\tau^2)$$

The substitution
$$ v  = (1-u)^{-1} - \ln|u-1|
$$
brings the rate of growth to
$$
\frac{dv}{dt} = 1 + \gamma (1+\psi)^2
$$
where $\psi$ satisfies $$\psi + \ln|\psi| = v$$
and asymptotically
$$
\frac{dv}{dt} = 1 + \gamma v^2
$$

The initial asymptotics in LS theory corresponds to $\gamma =
0$ and then asymptotically
$$
v = \tau+ const
$$

The  initial  asymptotics in OK theory corresponds to $\gamma = 1/(4
\tau^2)$. Then
asymptotically
$$v \rightarrow 2 \tau
$$

\section{Correction for the distributions}

The distribution function can be in frames of LS and
OK theories found from
$$
f(u,\tau) = \xi(C(u,\gamma)) \frac{\partial C}{\partial u}
$$
where $C$ is the integral of the law of growth.

Concrete calculations will give the following results
\begin{itemize}
\item
In LS theory
$$v - \tau = const
$$ or
$$
\exp(v-\tau) = const
$$
i.e.
$$
C_{LS} = \frac{1}{1-u} \exp(-\frac{1}{1-u} - \tau)
$$
is the integral of evolution
Then $\xi$ is
$$
\xi \sim C_{LS}^2 \ \ at \ \ C_{LS}>0, \ \ \ \ \
\xi \sim 0 \ \ at \ \ C_{LS}<0
$$
Then the spectrum is
$$
\phi_{LS} = \frac{2u}{(1-u)^4} \exp(-\frac{2u}{1-u})
$$
\item
In OK theory
$$v - 2  \tau = const
$$ or
$$
\exp(v- 2 \tau) = const
$$
i.e.
$$
C_{OK} = \frac{1}{1-u} \exp(-\frac{1}{1-u} -  2 \tau)
$$
is the integral of evolution
Then $\xi$ is
$$
\xi \sim C_{OK} \ \ at \ \ C_{OK}>0, \ \ \ \ \
\xi \sim 0 \ \ at \ \ C_{OK}<0
$$
Then the spectrum is
$$
\phi_{OK} = \frac{u}{(1-u)^3} \exp(-\frac{u}{1-u})
$$
and it essentially differs from LS theory.
\end{itemize}

\section{Further corrections}

The essential difference of spectrums is the striking
feature of the LS theory. But it is rather
easy to see that such a feature will be natural for all
further corrections.

Suppose that the integral of evolution at the previous
step is established. Then we can reformulate the rate of growth as
$$
\frac{du}{dt} = F(u) - \gamma u
$$
where $F$ is a known function.

Here we use for simplicity $t$ instead of $\tau$.

In the  zero approximation  (LS theory)
$$
F \sim  - u^{-1}+1/2
$$
in the first approximation\footnote{With a renormalization} (OK theory)
$$
F=(u-1)^2/u
$$
etc.

Let $w$ be
the solution of the equation with $\gamma=0$, i.e.
$$
\int\frac{du}{F(u)} = t + const
$$
Then at the previous step
$$
\frac{dw}{dt} = 1
$$
At the current step
$$
\frac{dw}{dt} = 1+ \gamma u(w)
$$
So, $w$ is a straight analog of $v$.

There can be two situations:
\begin{enumerate}
\item
 Asymptotically $w
\rightarrow t$.
\item
Asymptotically $w
\rightarrow l(t) \neq t$.
\end{enumerate}

In the first situation the spectrum remains absolutely the
previous one. There is absolutely no corrections.

In the second situation the correction is essential. Instead of
$w-t$ as the integral of evolution one has to use
$w-l(t)$. It means that instead of $w$ one has to use
$l^{\{-1\}}(w)$ where $l^{\{-1\}}$ is the inverse function.

The difference in spectrums is striking.

So, the alternative is to have no corrections or the
striking corrections. This is the consequence of the use of
Lifshic-Slezov variables and a certain disadvantage of LS and OK approaches.

\section{Application of asymptotics}

The ideology of LS and OK asymptotic analysis is one  and
the same:
\begin{itemize}
\item
The asymptotic of some given function
of the supersaturation in the class of powers is
prescribed.
\item
The balance condition at asymptotics leads to the
determination of the supersaturation as the function of
time.
\item
The form of the size spectrum is
reconstructed on the base of the given supersaturation.
\end{itemize}

Meanwhile, under the regular growth of the embryos the
natural sequence of actions is the following
\begin{itemize}
\item
The size spectrum is given from initial conditions
\item
The dissolution of the size spectrum together with the
balance equation determines the behavior of the
supersaturation
\item
The tail of the spectrum at big sizes determines the
asymptotics of the supersaturation and of coefficients in
the regular law of the droplets motion.
\end{itemize}

We see that from the last point of  view the LS and OK
theories are  inconsistent. Then why they
correspond to the asymptotic behavior found from
experiment?

At  first one has to mention that the accuracy of
experimental measurements is not high: even the
striking difference
between LS and OK drops out of experiments.
Hence, we have to conclude that the
experiment gives only the approximate form of the size
spectrum with maximum and two different rather short wings.

Now we shall see that the form of the size spectrum
obtained in LS and OK theories is rather typical at least
approximately.

The tail of the Gaussian can well approximated by
$$
\xi_{as} = \exp(- const * \rho)
$$
with some constant $const$. We shall call this constant
$\Lambda$ and suppose it to be a slow function of $\rho$ to
have
$$
\Lambda(\rho+ \Lambda^{-1}) - \Lambda(\rho) \ll
\Lambda(\rho)
$$
Then the exponential approximation will be valid at least
at the essential part of the size spectrum.

These transformations are  absolutely adequate to the
standard approach of the steepest descent method.

The same approximation can be established in $u$-scale. The
constant can simply cancelled by renormalization. Then
$$\xi_{as} \sim  \exp(-u)$$

Now we  shall approximate $\tau(u)$ for two values $u_1$
and $u_2$  at the essential part of the size spectrum. The
simplest approximation is the following
$$
\tau(u_1) - \tau(u_2) = \frac{d \tau(u)}{du} (u_1 - u_2)
$$
which is the linear connection. Then
the exponential form of the spectrum over $\tau(u)$ will be
conserved
$$
\xi_{as} \sim \exp(-\tau(u))
$$

Since $\xi$ has to be the function of $\tau - \tau(u)$ we
come to
$$
\xi = \exp(\tau - \tau(u))
$$
which lies in the base of LS and OK reconstruction of the
size spectrum.

To get the spectrum one has to divide $\xi$ on $v_u$. The
last value is determined by the regular law of growth.
Certainly, one can not
guarantee that the parameter $\gamma$  corresponds to $r=0$.
When it really corresponds to $r=0$ there will be LS
asymptotic.
When $\gamma$ corresponds to some small $r$  less than zero
 then there will be an extremum associated with the minimum
 of $v_u$.

One can see that since $r$ is rather small in the absolute
value there is a maximum of $\phi$ near $u_0$. One has to
note that in the experiment there is no evident way to
determine the critical size. Then there is no way to
determine the position of the size spectrum, but only it's
form.

Since $\Lambda$ is not a true constant one can say about
the quasistationary solution with a smooth variation of
parameters and a form of the spectrum.

These considerations evoke the ideas associated with the
square approximation of the rate of growth $du/d\tau$ near
the maximum which were presented in \cite{book1}.

\section{Direct determination of $\gamma$.}

Although the precise asymptotic  can not be directly used
to determine the evolution of the system one can suggest
another approach.

Now we return to the description of the evolution.

The rate of growth can be written as
$$
\frac{du}{d\tau} =
\hat{F}(u, \gamma(\tau))
$$
The integration of this equation gives $\tilde{\tau}(u)$.
Symbolically this can be written as
$$
\tilde{\tau}(u) = \int \frac{du}{v_u}
$$
where
$$
v_u = \frac{du}{d\tau}
$$
The problem of  integration of the rate of growth is very
important and analytically it is difficult to do, but
here we suppose that it is fulfilled.

The balance equation has to written as
$$
G_{total} = const
$$ It  is possible to neglect the surplace mother
phase. It can be reduced to
$$
\exp(\tau) \int_0^{u_{lim}} u^3 \phi(u,\tau)
du = const
$$
where $\phi$ is the distribution function, $u_{lim}$ is the maximal size.

The regular law of growth leads to
$$
\phi(u, \tau) =
\frac{
\xi(\tau - \tilde{\tau}(u)) }{ -v_u}
$$
Then the balance equation will be
$$
\exp(\tau) \int_0^{u_{lim}}  \frac{u^3\xi(\tau - \tilde{\tau}(u)) }{ -v_u
}
du = const
$$
But the last equation is not an equation on the function $\xi$
as in the LS and OK theories but an equation on $\tilde{\gamma}$.
The function $\xi$ is known from the initial conditions for
this equation.
This is the crucial difference between our theory and LS or
OK theory.
The dependence $\tilde{\gamma}(\tau)$ stands in $v_u$ and in
$\tilde{\tau}(u)$. Moreover in $\tilde{\tau}(u)$ it is at
least the dependence on $\tilde{\gamma}$ in all preceding moments
of time. So, the last equation is very complex and one gas
to suggest some methods of the approximate solution.

\section{Approximate solution}

When $\tilde{\gamma}$ is really smaller than $\tilde{\gamma}_0$ corresponding
to $r=0$ it is possible to see that some approximations for
$\tilde{\gamma}$ produces approximations for evolution.
When the main influence is ensured by the supercritical
tail then the condition
$$
G_{tail} = const
$$
is rather productive.

One can approximately say that in the region $\rho <
\rho_b$ the rate of growth is very small.

Another possibility is to use for $du/d\tau$ an
approximation which can be integrated analytically.
Namely $du/d\tau$ has to be used because for $u$ the
boundaries of the nearcritical region
and supercritical region will be fixed. For
$$
\frac{du}{d\tau} = (1-u^{-1}) - \tilde{\gamma} u
$$
we use the following constructions:
\begin{itemize}
\item
We give definitions
$$
F_1 = (1-u^{-1})
$$
$$
F_2 = r
$$
$$
F_3 = 1 - \tilde{\gamma} u
$$
\item
The first interval will be $[u_2, \infty[$. The value of
$u_2$ is the root of equation
$$
F_2=F_3
$$
At the first interval one can approximately write
$$
\frac{du}{d\tau} = F_3
$$
The last equation can be integrated analytically
\item
The second interval is
$[u_1,u_2]$
where
$u_1$ is the root of equation
$$
F_2=F_1
$$
At the second interval one can approximately write
$$
\frac{du}{d\tau} = F_2
$$
The last equation can be integrated analytically
\item
The third interval is
$[0,u_1]$.
At the third interval one can approximately write
$$
\frac{du}{d\tau} = F_1
$$
The last equation can be integrated analytically
\end{itemize}

Summarizing one can approximately write
$$
\frac{du}{d\tau} = \bar{F}(u,\tilde{\gamma})
$$
with an analytical solution.
Then we know $\bar{\tau}(u)$ and can write a closed
equation on $\tilde{\gamma}$.

This approach will be called as the "regular-regular
model".

In the  general situation allowing LS asymptotics
$$
du/d\tau = F_1 - \tilde{\gamma} u
$$
and the procedure remains the same.

Another possibility is  the \underline{quasistationary
approximation}.
Here $\tilde{\gamma}$ is supposed to be a local constant and the law of
growth is integrated
$$
\tau(u) = \int\frac{du}{F_1 - \tilde{\gamma} u}
$$
Then we know $\tau(u)$ and $\xi(\tau - \tau(u))$ as function
of $u,\tau$. Then the balance equation is the closed equation on $\tilde{\gamma}$.

Define as $\tau_{total}$ the time of dissolution from the
size $u_m$, where $du/d\tau$ attains maximum, up to zero.
It is clear that the necessary condition for the
applicability of the quasistationary approximation is
$$
\frac{|\tau_{total}(\tilde{\gamma}(\tau)) -
\tau_{total}(\tilde{\gamma}(\tau+\tau_{total}(\tilde{\gamma}(\tau))))|
}{\tau_{total}(\tilde{\gamma}(\tau))} \ll 1
$$

The difference between LS and OK theories symbolizes that
at the asymptotics $r=0$ the quasistationary approximation
can be hardly applied at the very end of the process. But
it can be fruitful at some earlier periods.

\section{Combination of approaches}

Here the regular-regular model and the diffusion-regular
models of solution were presented. Now it is necessary to
decide what model has to be used.

The main object will be the support of spectrum. We shall
define the essential support of the spectrum by the
following way:
\begin{itemize}
\item
For $G_{total}$ we define the subintegral function
$g$ by
$$
G_{total} = \int d \rho g_{\rho}
$$
or
$$
G_{total} = \int d u g_{u}
$$
\item
We define the maximum of $g$ and
$$\rho_{max} =
\arg(max(g))$$
$$u_{max} = \arg(max(g))$$
\item
Then we determine the region of support
by
$$
\int_0^{u_l} du g_u = \delta/2
$$
$$
\int_{u_r}^\infty du g_u = \delta/2
$$
or by
$$
\int_0^{\rho_l} d\rho g_\rho = \delta/2
$$
$$
\int_{\rho_r}^\infty d\rho g_\rho = \delta/2
$$
with some small parameter $\delta$.

This region can be also determined by conditions
$$
g(u_l) = g(u_r) = \delta g(u_{max})
$$

\end{itemize}

If the essential support belongs to the near-critical
region then we have to use the diffusion-regular model.
In the opposite situation we have to the regular-regular
model.

The impulse regime contains the sequential application of
the diffusion-regular model and the regular-regular
model. Here the evolution during the period described by
the regular-regular model occurs in a very simple manner -
the essentially supercritical droplets determine the
behavior of the supersaturation. So, there is no need to
use the complete formalism of the regular-regular model.

In the regime of adjusted tail one has to use the
regular-regular model which can not be simplified.

To see the times when the regular-regular model will be no
longer valid it is possible to apply the diffusion operator
to the solution obtained in the regular-regular model. When
the difference between this solution and the result of the
application of diffusion operator will be essential it means
that it is necessary to introduce the diffusion
corrections.

The presence of diffusion corrections will be very
essential at the tail of the size spectrum and namely this
tail will govern the evolution later. This tail can be
hardly described by the diffusion equation because there is
no statistics - only few droplets of big sizes determine
the evolution.

Ordinary, the process of nucleation results in existence of
several big droplets with stochastic sizes and namely their
competition in growth determines the latest stage of the
process. The sizes of these droplets are rather arbitrary.
So, the kinetics of interaction will be also unstable and
arbitrary.
This effect forms a matter of a special
publication.

Here it is necessary only to add that the
number of droplets strongly diminishes in time and to see
the impulse regime and later the regime of adjusted tail it
is  necessary to have a nucleating system of huge  sizes.


\begin{thebibliography}{99}

\bibitem{ls}
 Lifshitz I.M., Slyozov V.V. The kinetics of precipitation
from supersaturated solid soluitons -
J.Phys.Chem.Solids, 1961, v.19, N 1/2,
p.35

\bibitem{kukos}
Kukushkin S.A., Osipov A.V. Kinetics of the first order phase
transitions at the asymptotic stage, Journal of
Experimental and Theoretical Physics, vol. 113, issue 6,
p.2193-2208 (1998)

\bibitem{kukos1}
Kukushkin S., Osipov A,
J.Chem.Phys. vol. 107 p 3247-3252  (1997)



\bibitem{kuni}
Kuni F.M., Grinin A.P., Kabanov A. S.
Kolloidnui journal vol 46, p 440 (1984)


\bibitem{preprint}
Kurasov V. General trends of the late period
of evolution in the quasichemical
model of nucleation, Preprint  arXiv.org chemical physics 0607768
(2006)


\bibitem{book1}
Kurasov V. B., Universality in kinetics
of the first order phase transitions,
 Chemistry
Research Institute of Sankt-Petersburg
State University, St.Peterburg, 1997,
400 p.


\bibitem{book2}
Kurasov V.B., Development of the
universality conception in the first
order phase transitions,
 Chemistry
Research Institute of Sankt-Petersburg
State University, St.Peterburg, 1998, 125
p.


\end{thebibliography}
\end{document}